\documentclass{elsart}
\usepackage{graphics}
\usepackage{graphicx}
\usepackage{epsfig}
\usepackage{amssymb,amsmath}
\usepackage[colorlinks]{hyperref}

\newtheorem{lemma}{Lemma}[section]
\newtheorem{corollary}{Corollary}[section]
\newtheorem{theorem}{Theorem}[section]

\newtheorem{definition}{Definition}[section]

\begin{document}
\linespread{1}
\begin{frontmatter}
%\bibliographystyle{fr-plain}
\title{A wavelet uncertainty principle in quantum calculus}
\author{Sabrine Arfaoui\corauthref{Label1}}
\address{Laboratory of Algebra, Number Theory and Nonlinear Analysis, LR18ES15, Department of Mathematics, Faculty of Sciences, 5000 Monastir, Tunisia.\\
\& Department of Mathematics, Faculty of Sciences, University of Tabuk, Saudi Arabia.} \ead{sabrine.arfaoui@issatm.rnu.tn}
\corauth[Label1]{Corresponding author.}
\author{Maryam G. Alshehri}
\address{Laboratory of Algebra, Number Theory and Nonlinear Analysis, LR18ES15, Department of Mathematics, Faculty of Sciences, 5000 Monastir, Tunisia.\\
\& Department of Mathematics, Faculty of Sciences, University of Tabuk, Saudi Arabia.} \ead{mgalshehri@ut.edu.sa}
\author{Anouar Ben Mabrouk}
\address{Department of Mathematics, Higher Institute of Applied Mathematics and Computer Science, Street of Assad Ibn Alfourat, 3100 Kairouan, Tunisia.\\
\& Laboratory of Algebra, Number Theory and Nonlinear Analysis, LR18ES15, Department of Mathematics, Faculty of Sciences, 5000 Monastir, Tunisia.\\
\& Department of Mathematics, Faculty of Sciences, University of Tabuk, Saudi Arabia.}
\ead{anouar.benmabrouk@fsm.rnu.tn}
%
%%%%%%%%%%%%%%%%%%%%%%%%%%%%%%%%%%%%%%%%%%%%%%%%%%%%%%%%%%%%%%%%%%%%%%%%%%%%%%%%%%%%%%%%%%%%%%%%%%%%%%%%%%%%%%%%%%%%%%%%%%%%%%%%%%%%%%%%%%%%%%%%
%%%%%%%%%%%%%%%%%%%%%%%%%%%%%%%%%%%%%%%%%%%%%%%%%%%%%%%%%%%%%%%%%%%%%%%%%%%%%%%%%%%%%%%%%%%%%%%%%%%%%%%%%%%%%%%%%%%%%%%%%%%%%%%%%%%%%%%%%%%%%%%%
%%%%%%%%%%%%%%%%%%%%%%%%%%%%%%%%%%%%%%%%%%%%%%%%%%%%%%%%%%%%%%%%%%%%%%%%%%%%%%%%%%%%%%%%%%%%%%%%%%%%%%%%%%%%%%%%%%%%%%%%%%%%%%%%%%%%%%%%%%%%%%%%
%%%%%%%%%%%%%%%%%%%%%%%%%%%%%%%%%%%%%%%%%%%%%%%%%%%%%%%%%%%%%%%%%%%%%%%%%%%%%%%%%%%%%%%%%%%%%%%%%%%%%%%%%%%%%%%%%%%%%%%%%%%%%%%%%%%%%%%%%%%%%%%%
\begin{abstract}
In the present paper, a new uncertainty principle is derived for the generalized $q$-Bessel wavelet transform studied earlier in \cite{Rezguietal}. In this paper, an uncertainty principle associated with wavelet transforms in the $q$-calculus framework has been established. A two-parameters extension of the classical Bessel operator is applied to generate a wavelet function which is exploited next to explore a wavelet uncertainty principle already in the $q$-calculus framework. 
\end{abstract}
\end{frontmatter}
%\tableofcontents
%
\section{Introduction}
Wavelet analysis also called Wavelet theory has attracted much attention recently in different domains such as mathematics, quantum physics, electrical engineering, seismic geology.  Interchanges between these fields during the last ten years have led to many new wavelet applications in astronomy, acoustics, nuclear engineering, sub-band coding, signal and image processing, neurophysiology, music, magnetic resonance imaging, speech discrimination, optics, fractals, turbulence, earthquake-prediction, radar, human vision, and pure mathematics applications such as solving partial differential equations.

Like Fourier analysis wavelet analysis deals with expansions of functions in terms of a set of basis functions. Unlike Fourier analysis, wavelet analysis expands functions not in terms of trigonometric polynomials but in terms of wavelets which are a family of functions obtained from one function known as the mother wavelet, by translations and dilations. This tool permits the representation of $L^2$-functions in a basis well localized in time and in frequency. The fundamental idea behind wavelets is to analyze according to scale. Indeed, some researchers in
the wavelet field feel that, by using wavelets, one is adopting a whole new mindset or perspective in processing data. 

Hence, wavelets are special functions characterized by special properties that may not be satisfied by other functions. 

Bessel functions form an important class of special functions and are applied almost everywhere in mathematical physics. The Bessel function was the result of Bessels study of a problem of Kepler for determining the motion of three bodies moving under mutual gravitation.
In 1824, he incorporated Bessel functions in a study of planetary perturbations where the Bessel functions appear as coefficients in a series expansion of the indirect perturbation of a planet, that is the motion of the Sun caused by the perturbing body. It was likely
Lagrange’s work on elliptical orbits that first suggested to Bessel to work on the Bessel functions. They are also known as cylindrical functions, or cylindrical harmonics, because of their strong link to the solutions of the Laplace equation in cylindrical coordinates. 

One of the interesting fields of extensions of Fourier and wavelet analyses is the so-called $q$-theory which is an important sub-field in harmonic analysis and which provides some discrete and/or some refinement of continuous harmonic analysis in sub-spaces such as $\mathbb{R}_q$ composed of the discrete grid $\pm q^n$, $n\in\mathbb{Z}$, $q\in(0,1)$. Recall that for all $x\in\mathbb{R}^*$ there exists a unique $n\in\mathbb{Z}$ such that $q^{n+1}<|x|\leq q^n$ which guarantees some density of the set $\mathbb{R}_q$ in $\mathbb{R}$.

Many special functions have been shown to admit generalizations to a base $q$, and are usually reported as $q$-special functions. Interest in such $q$-functions is motivated by the recent and increasing relevance of $q$-analysis in exactly solvable models in statistical mechanics. Like ordinary special functions, $q$-analogues satisfy second order $q$-differential equations and various identities of recurrence relations. Basic analogues of Bessel function have been introduced by Jackson and Swarttow as $q$-generalizations of the power series expansions.

In the present context, We aim to apply the generalized $q$-Bessel and associated $q$-wavelets introduced in the context of $q$-theory to develop new uncertainty principle based on wavelets in the framework of $q$ (quantum)-calculus. We serve of the new $q$-wavelets developed recently in \cite{Rezguietal} to develop a new $q$-wavelet uncertainty principle. 

The present paper is organized as follows. In section 2, a brief review focusing on the last developments in $q$-wavelets' theory is developed. Section 3 is concerned to a review on the uncertainty principle. In section 4, we present our main results. We precisely introduce the new generalized $q$-wavelet uncertainty principle. Section 5 is concerned with the proofs of our main results. Finally section 6 presents our conclusion.
\section{$q$-calculus toolkit}
The present section aims to recall the basic tools in $q$-calculus. For $0<q<1$, denote
$$
\mathbb{R}_{q}=\{\pm q^{n},\;n\in\mathbb{Z}\},\;
\widetilde{\mathbb{R}}_{q}=\mathbb{R}_{q}\cup \{0\},\;
\mathbb{R}_{q}^+=\{q^{n},\,n\in\mathbb{Z}\}
\;\hbox{and}\;
\widetilde{\mathbb{R}}_{q}^{+}=\mathbb{R}_{q}^{+}\cup \{0\}.
$$
On $\widetilde{\mathbb{R}}_{q}^{+}$, the $q$-Jackson integrals on $[a,b]$ ($a\leq b$ in $\widetilde{\mathbb{R}}_q^+$) is defined by
$$
\displaystyle\int_{a}^{b}f(x)d_{q}x=(1-q)\sum_{n\in\mathbb{Z}}q^n(bf(bq^{n})-af(aq^n))
$$
and for $[0,+\infty)$ it is defined by
$$
\displaystyle\int_{0}^{\infty}f(x)d_{q}x=(1-q)\sum_{n\in\mathbb{Z}}f(q^{n})\,q^{n}
$$
provided that the sums converge absolutely. On $[a,+\infty)$, we have to apply an analogue of Chasle's rule and thus get
$$
\displaystyle\int_{a}^{+\infty}f(x)d_{q}x=\displaystyle\int_{0}^{+\infty}f(x)d_{q}x-\displaystyle\int_{0}^{a}f(x)d_{q}x.
$$
(See \cite{Slim}, \cite{Ahmed}). This leads next to define the functional space
$$
\mathcal{L}_{q,p,v}(\widetilde{\mathbb{R}}_{q})=\{f:\mathbb{\widetilde{R}}_q\rightarrow\mathbb{C},\;\mbox{even}\;;\;\,\|f\|_{q,p,v}<\infty\},
$$
where $\|.\|_{q,p,v}$ is the norm defined analogously by
$$
\|f\|_{q,p,v}=\left[\displaystyle\int_{0}^{\infty}|f(x)|^{p}\,x^{2|v|+1}\,d_{q}x\right]^{\frac{1}{p}}.
$$
where $v$ is a fixed real parameter such that $2v>-1$. Denote next, $C_{q}^{0}(\widetilde{\mathbb{R}}_{q}^{+})$ the space of functions defined on $\widetilde{\mathbb{R}}_{q}^{+}$, continuous at $0$ and vanishing at $+\infty$, equipped with uniform norm
$$
\|f\|_{q,\infty}=\sup_{x\in\widetilde{\mathbb{R}}_{q}^{+}}|f(x)|<\infty.
$$
Finally, $C_{q}^{b}(\widetilde{\mathbb{R}}_{q}^{+})$ designates the space of functions that are continuous at $0$ and bounded on $\widetilde{\mathbb{R}}_{q}^{+}$. Recall here that $q^n\rightarrow\infty$ as $n\rightarrow-\infty$ in $\mathbb{Z}$. The $q$-derivative of a function $f\in\mathcal{L}_{q,p,v}(\widetilde{\mathbb{R}}_{q}^{+})$ is defined by
$$
D_{q}f(x)=\left\{\begin{array}{lll}
\dfrac{f(x)-f(qx)}{(1-q)x},\;\;&\;x\neq 0\\
f'(0)\;,& else.
\end{array}\right.
$$	
In $q$-theory, we may also have an analogue of the change variables theorem. Particularly,
\begin{equation}\label{changeVariables}
\displaystyle\int_{0}^{\infty}f(t)t^{2|v|+1}\,d_{q}t=
\displaystyle\frac{1}{a^{2v+2}}\displaystyle\int_{0}^{\infty}f(\displaystyle\frac{x}{a})x^{2v+1}\,d_{q}x.
\end{equation}
The $q$-shifted factorials are defined by
$$
(a,q)_{0}=1,\;\;\;(a,q)_{n}=\prod_{k=0}^{n-1}(1-aq^k),\;\;\;(a,q)_{\infty}=\prod_{k=0}^{+\infty}(1-aq^k).
$$
\section{A literature review on the uncertainty principle}

The uncertainty principle has been introduced at the begining of the early 20th years of the last century by Heisenberg, and it toke consequently the name of Heisenberg uncertainty principle (\cite{Heisenberg1927,Heisenberg1985}. Mathematically speaking, the uncertainty principle formula states that a non-zero function and its Fourier transform cannot be both sharply localized. It is thus a consequence of Fourier transform. It mathematically explains the functional infinite dimensionality property (See \cite{Das}). In physics, the uncertainty principle states that the determination of positions by performing measurement on the system disturbs it sufficiently to make the determination of momentum imprecise and vice-versa. In quantum mechanics, it may be explained by the existence of a fundamental limit to the precision with which it is possible to simultaneously know the position and the momentum of the particle. The uncertainty principle interacts noadays with many fields such as pure mathematics, physics, engineering, communication, quantum mechanics. 

The literature on the uncertainty principle is sure wide, and studies investigating such concept are multidisciplinary. Exploitations and also generalizations of the uncertainty principle have been developed. The recent variants are those based on wavelet theory. Our present work lies in theis last topic, and constitutes a contribution in the uncertainty principle based on Bessel theory, wavelet theory and quantum calculus. 

In \cite{Amri-Rachdi-1} established a Pitt's and Beckner logarithmic uncertainty inequalities using Fourier transform. In \cite{Amri-Rachdi-2,Hleilietal}, Fourier-Riemann–Liouville operator has been applied to derive an entropy based uncertainty principle and a Heisenberg–Pauli–Weyl inequality. Rachdi and collaborators derived two types of uncertainty principle such as Heisenberg-Pauli-Weyl and Beurling-H\"ormander in \cite{Msehli-Rachdi-1,Msehli-Rachdi-2}, by applying Fourier and spherical mean operators. 

In wavelet theory framework, Rachdi and Mehrzi have derived a Heisenberg's uncertainty principle by combining the continuous wavelet transform with spherical mean operators \cite{Rachdi-Meherzi}. Extensions of this result have been developed in \cite{Rachdi-Amri-Hammami,Rachdi-Herch}. A conctinuous shearlet uncertainty principle has been investigated in \cite{Dahkleteal} provided with the computation of some minimizers of such uncertainty principle.

As an extension of wavelet theory, and harmonic analysis in general, Clifford wavelet analysis has taken place especially in the last decades. The Heisenberg uncertainty principle and its different variants and extensions have been investigated in the large field of Clifford analysis. In \cite{ElHaouietal}, the authors investigated a quaternionic linear transform and combined it with quaternion Fourier transform to derive different uncertainty principles such as Heisenberg-Weyls, Hardys, Beurlings and logarithmic ones. El-Houi et al established in \cite{ElHaoui-Fahlaoui} some Clifford uncertainty inequalities such as Hausdorf-Young, and Donoh-Stark uncertainty principles.

Feichtinger and Grochenig applied the Heisenberg uncertainty principle dut to Gabor wavelets and short time Fourier tranform for the control of signals \cite{Feichtinger-Grochenig}. Hitzer et al introduced in \cite{Hitzer1} a general formula for the uncertainty principle based on the local Clifford geometric algebra wavelet transform. In \cite{Hitzer2} the same author derived a new directional uncertainty principle for quaternion valued functions based on Clifford quaternion Fourier transform. In \cite{Hitzer-Tachibana}, Clifford-Fourier transform has been applied to establish an uncertainty principle in the case of Clifford wavelets. Such a principle has been shown to be useful in signal processing. See also \cite{Hitzer-Mawardi-1,Mawardi-Ryuichi,Mawardi-Ryuichi-1,Mawardi-Ryuichi-2,Mawardi-Hitzer-1,Mawardi-Hitzer-2,Mawardi-Hitzer-3,Mawardi-Hitzer-4,Mawardi-Hitzer-Hayashi-Ashino,Mawardi-Ashino-Vaillancourt} for the same team of researchers and for eventual extensions of their works.

Ma and Zhao applied quaternion ridgelet transform and curvelet transform to derive an associated uncertainty principles \cite{Ma-Zhao}. In \cite{Kouetal} a quaternionic linear canonical transform has been applied to derived an uncertainty principle for vector-valued signals. The authors showed that Gaussian quaternion signals are the only minimizers for the uncertainty principle formula.

In \cite{Mejjaolietal}, a hybrid method combing continuous wavelet transform and spherical mean operators has been used for both Donoho-Stark and Benedicks-type uncertainty principles. Mejjaoli et al proved in \cite{Mejjaolietal1} an inversion formula due to the Dunkl Gabor transform. Many versions of the uncertainty principle has been shown such as Heisenberg, Donoho–Stark's, local Cowling–Price's, and Faris-Price's uncertainty principles.

In \cite{Yangetal1} a stronger uncertainty principles has been derived in terms of covariance and absolute covariance based on Fourier transform for vector-valued signals. Yang and Kou in \cite{Yang-Kou} extended the uncertainty principle to the case of hypercomplex signals by means of the linear canonical transform. Besides, they proved that Gaussian signals are the only minimizers. 

In $q$-theory, uncertainty principle has been the subject of several studies, in both Fourier and wavelet frameworks. In \cite{Fitouhietal1}, the authors developed a general $q$-Heisenberg uncertainty principle for the $q$-Fourier-cosine/sine and the $q$-Bessel-Fourier transform. In \cite{Fitouhietal2}, a heisenberg-Weyl type uncertainty principle has been shown for a $q$-Dunkl transform. Besides, a sharp $q$-Bessel-Dunkl uncertainty principle has been proved in \cite{Fitouhietal3} generalizing the case of basic Bessel transform. In \cite{Fitouhietal4}, the $q$-Dunkl Transform on the Real Line has been applied to derive an $L_p$-version of the Hardy Uncertainty Principle. 

In \cite{Hleili}, Weinstein operator has been applied to introduce a new wavelet transform, which is used by the next to derive a Heisenberg uncertainty principle. Nemri in \cite{Nemri} investigated an extension of the Donoho-Stark's uncertainty principle for a special class of Fourier multiplier operators. Ogawa et al (\cite{Ogawa-Seraku}) applied the Boltzmann entropy functional to derive a Heisenberg-type uncertainty principle. Saoudi et al (\cite{Saoudietal}) proved a Heisenberg-Pauli-Weyl uncertainty principle, and Donho-Stark's uncertainty principle associated to the Weinstein $L_2$-multiplier operators.

The uncertainty principle in its classical or original form may be explained by the following result, which will be the starting step to our extension in the present work. 
\begin{theorem}\label{uncertainty-cas-general} (\cite{Banouh1,Banouh2,Weyl1950}) Let $A$ and $B$ be two self-adjoint operators on a Hilbert space $X$ with domains $\mathcal{D}(A)$ and $\mathcal{D}(B)$ respectively and consider their commutator $\left[A,B\right]=AB-BA$. Then
\begin{equation}
\left\|Af\right\|_{2}\left\|Bf\right\|_{2}\geq\displaystyle\frac{1}{2}\left|<\left[A,B\right]f,f>\right|,\forall f\in\mathcal{D}(\left[A,B\right]).
\end{equation}
\end{theorem}
Consider next the special case 
$$
A_{k}f(\underline{x})=x_{k}f(\underline{x})\;\;\hbox{and}\;\;
B_{k}f(\underline{x})=\partial_{x_{k}}f(\underline{x}),\;k=1,2,\cdots,n.
$$
By applying Theorem \ref{uncertainty-cas-general} above, we get the following result.
\begin{corollary}\label{cor1}(\cite{Banouh1,Banouh2}) 
\[
\left\|A_{k}f\right\|_2\left\|B_{k}f\right|_2\geq\frac{1}{2}\left|<\left[A_{k},B_{k}\right]f,f>\right|.
\]
Furthermore,
\begin{equation}\label{uncerclifford-fourier}
\|x_{k}f\|_2\|\xi_{k}\widehat{f}\|_2\geq\frac{1}{2}\|f\|_2^{2}.
\end{equation}
\end{corollary}
For more backgrounds on the uncertainty principle, its variants, Fourier and wavelet transforms on the Euclidean space $\mathbb{R}^n$ the readers may be referred also to \cite{Jorgensen}, \cite{Nagata}, \cite{Sen}, \cite{Stabnikov}, \cite{Yang-Kou}.
\section{Main results}
In \cite{Rezguietal}, a general two-parameters $q$-wavelet analysis has been developed generalizing thus quasi all the previous versions of $q$-wavelets. The idea is based on \cite{Manel}, where a two-parameters $q$-theory leading to a two-parameters $q$-Bessel function has been introduced provided with suitable associated Fourier transform. The authors in \cite{Rezguietal} exploited such a $q$-version of the Bessel function to construct $q$-wavelets and develop an associated $q$-wavelet analysis. The parameter $v$ in the classical $q$-Bessel theory is replaced by a couple of parameters $v=(\alpha,\beta)\in\mathbb{R}^2$, $\alpha+\beta>-1$, for which a modified functional space is associated with suitable integrating measure. For $1\leq p<\infty$, we denote
$$
\mathcal{L}_{q,p,v}(\widetilde{\mathbb{R}}_{q}^{+})=\left\{f: \,\|f\|_{q,p,v}=\left[\displaystyle\int_{0}^{\infty}|f(x)|^{p}\,x^{2|v|+1}\,d_{q}x\right]^{\frac{1}{p}}<\infty\right\}.
$$
where $|v|=\alpha+\beta$. The generalized $q$-Bessel operator is defined by
$$
\widetilde{\Delta}_{q,\,v}f(x)=\dfrac{f(q^{-1}x)-(q^{2\alpha}+q^{2\beta})f(x)+q^{2\alpha+2\beta}f(qx)}{x^{2}},\;\;\;\;\forall x\neq 0.
$$
The normalized $q$-Bessel function is given by
$$
j_{\alpha}(x,q^{2})=\displaystyle\sum_{n\geq 0}(-1)^{n}\dfrac{q^{n(n+1)}}{(q^{2\alpha+2},q^{2})_{n}\,(q^{2},q^{2})_{n}}\,x^{2n}.
$$
We denote finally $\widetilde{j}_{q,v}(x,q^{2})$ the modified version of $q$-Bessel function is
\begin{equation}\label{0.32}
\widetilde{j}_{q,v}(x,q^{2})=x^{-2\beta}\,j_{\alpha-\beta}(q^{-\beta}x,q^{2}).
\end{equation}
More details may be found in \cite{Manel,Rezguietal}. The generalized $q$-Bessel Fourier transform $\mathcal{F}_{q,v}$, is defined by
\begin{equation}\label{0.46}
\mathcal{F}_{q,v}f(x)=c_{q,v}\displaystyle\int_{0}^{\infty}\,f(t)\,\widetilde{j}_{q,v}(tx,q^{2})\,t^{2|v|+1}\,d_{q}t,\;\;\; \forall f\in\mathcal{L}_{q,p,v}(\mathbb{R}_{q}^{+}).
\end{equation}
where $c_{q,v}$ is a suitable constant. This has induced a generalized $q$-Bessel wavelet already as an even function $\Psi\in\mathcal{L}_{q,2,v}(\widetilde{\mathcal{R}}_{q})$ satisfying an analogue admissibility condition as for the existing cases with suitable and necessary modifications. The associated translation operator has been already defined in \cite{Manel} by
\begin{equation}\label{0.50}
T_{q,x}^{v}f(y)=c_{q,v}\displaystyle\int_{0}^{\infty}\mathcal{F}_{q,v}f(t)\,\widetilde{j}_{q,v}(yt,q^{2})\,\widetilde{j}_{q,v}(xt,q^{2})\,t^{2|v|+1}\,d_{q}t.
\end{equation}
\begin{definition}\label{Intro-5.19a}\cite{Rezguietal} A generalized $q$-Bessel wavelet is an even function $\Psi\in\mathcal{L}_{q,2,v}(\widetilde{\mathbb{R}}_{q}^{+})$ satisfying the following admissibility condition:
$$
C_{v,\Psi}=\displaystyle\int_{0}^{\infty}|\mathcal{F}_{q,v}\Psi(a)|^{2}\,\dfrac{d_{q}a}{a}<\infty.
$$
The continuous generalized $q$-Bessel wavelet transform of a function $f\in\mathcal{L}_{q,2,v}(\widetilde{\mathbb{R}}_{q}^{+})$ at the scale $a\in\mathbb{R}_{q}^{+}$ and the position $b\in \widetilde{\mathbb{R}}_{q}^{+}$ is defined by
$$
C_{q,\Psi}^{v}(f)(a,b)=c_{q,v}\,\displaystyle\int_{0}^{\infty}f(x)\,\overline{\Psi_{(a,b),v}}(x)\,x^{2|v|+1}\,d_{q}x,\;\forall a\in\mathbb{R}_{q}^{+},\;\forall b\in\widetilde{\mathbb{R}}_{q}^{+},
$$
where
$$
\Psi_{(a,b),v}(x)=\sqrt{a}\mathcal{T}_{q,b}^{v}(\Psi_{a})\qquad\hbox{and}\qquad\Psi_{a}(x)=\dfrac{1}{a^{2|v|+2}}\,\Psi(\dfrac{x}{a}).
$$
\end{definition}
In the present work, we will exploit such wavelets to prove a $q$-variant of the uncertainty principle relative to the $q$-wavelet transform developed in \cite{Rezguietal}. 

Let $\Psi$ be a generalized $q$-Bessel wavelet in  $\mathcal{L}_{q,2,v}(\widetilde{\mathbb{R}}_{q}^{+})$ and $f\in\mathcal{L}_{2,q,v}$. Denote for $\mathcal{R}$ and $\mathcal{S}$ the operators defined on $\mathcal{L}_{q,2,v}(\widetilde{\mathbb{R}}_{q}^{+})$ by
$$
\mathcal{R}f(a,b)=bC_{q,\Psi}^{v}(f)(a,b)\quad\mbox{and}\quad
\mathcal{S}f(t)=t\mathcal{F}_{q,v}f(t).
$$
Denote also $d_q(a,b)=b^{2|v|+1}\dfrac{d_{q}a \,d_{q}b}{a^{2}}$. 

We prove a general two-parameters $q$-wavelet uncertainty principle, we consider the general case of $q$-wavelets developed in \cite{Rezguietal}. The following $q$-wavelet uncertainty principle holds.
\begin{theorem}\label{theorem1} Bessel wavelet uncertainty principle 
\begin{align*}
\|f\|_{q,2,v}^2\leq\,K_{q,v}
\Biggl(\displaystyle\int_{0}^{\infty}\displaystyle\int_{0}^{\infty}\left|\mathcal{R}f(a,b)\right|^2d_{q}(a,b)\Biggr)\Biggl(\displaystyle\int_{0}^{\infty}\left|\mathcal{S}f(t)\right|^2\;t^{2|v|+1}d_{q}t\Biggr)
\end{align*}
for some constant $C_{q,v}>0$ depending on $q$ and $v=(\alpha,\beta)$.
\end{theorem}
\section{Proof of main results}
To prove this result we need the following lemma.
\begin{lemma}\label{MainResultLemma}
$$
\displaystyle\int_0^\infty\displaystyle\int_0^\infty\left|b\mathcal{F}_{q,v}C_{q,\Psi}^{v}(f)(a,b)\right|^2\frac{d_qad_qb}{a^2}=K_1^{\psi}\left\Vert\xi\mathcal{F}_{q,v}f(\xi)\right\Vert_{2}^{2}.
$$
\end{lemma}
\textit{Proof.} Let us express firstly the Fourier transform 
$\mathcal{F}_{q,v}\Psi_{a,b,v}$ by means of $\mathcal{F}_{q,v}\Psi$. So, denote for simplicity
$$
\Lambda_{q,v}(t,s,\xi,b)=\widetilde{j}_{q,v}(ts,q^{2})\widetilde{j}_{q,v}(bs,q^{2})\widetilde{j}_{q,v}(t\xi,q^{2}).
$$
Observe that 
$$
\begin{array}{lll}
\mathcal{F}_{q,v}\Psi_{a,b,v}(\xi)&=&
c_{q,v}\displaystyle\int_0^\infty\Psi_{a,b,v}(t)\widetilde{j}_{q,v}(t\xi,q^{2})\,t^{2|v|+1}\,d_{q}t\\
&=&c_{q,v}^2\sqrt{a}\displaystyle\int_0^\infty T_{q,b}^{v}\Psi_{a}(t)\widetilde{j}_{q,v}(t\xi,q^{2})\,t^{2|v|+1}\,d_{q}t\\
&=&c_{q,v}^2\sqrt{a}\displaystyle\int_0^\infty\displaystyle\int_0^\infty \mathcal{F}_{q,v}\Psi_{a}(s)\Lambda_{q,v}(t,s,\xi,b)\,(st)^{2|v|+1}d_qs\,d_{q}t\\
&=&c_{q,v}^2\sqrt{a}\displaystyle\int_0^\infty\displaystyle\int_0^\infty \mathcal{F}_{q,v}\Psi(as)\Lambda_{q,v}(t,s,\xi,b)(st)^{2|v|+1}d_qs\,d_{q}t\\
&=&\sqrt{a}\displaystyle\int_0^\infty \mathcal{F}_{q,v}\Psi(as)\delta_{q,v}(s,\xi)\widetilde{j}_{q,v}(bs,q^{2})s^{2|v|+1}d_qs\\
&=&\sqrt{a}\mathcal{F}_{q,v}\Psi(a\xi)\widetilde{j}_{q,v}(b\xi,q^{2}).
\end{array}
$$
As a consequence, we get
$$
C_{q,\Psi}^{v}(f)(a,b)=\sqrt{a}
\mathcal{F}_{q,v}^{-1}\left[\mathcal{F}_{q,v}\psi(a.)\mathcal{F}_{q,v}f(.)\right](b)
$$
and
$$
\mathcal{F}_{q,v}C_{q,\Psi}^{v}(f)(a,b)=\sqrt{a}\left[\mathcal{F}_{q,v}\psi(a.)\mathcal{F}_{q,v}f(.)\right](b).
$$
Therefore,
$$
\begin{array}{lll}
&\displaystyle\int_0^{\infty}\left|b\mathcal{F}_{q,v}C_{q,\Psi}^{v}(f)(a,b)\right|^{2}d_qb\\ 
&=\displaystyle\int_0^{\infty}\left|b\sqrt{a}\left[\mathcal{F}_{q,v}\psi(a.)\mathcal{F}_{q,v}f(.)\right](b)\right|^{2}d_qb\\
&=\displaystyle\int_0^{\infty}\left|\sqrt{a}\mathcal{F}_{q,v}\psi(a.)\right|^2
\left|b\mathcal{F}_{q,v}f\right|^2d_qb.
\end{array}
$$
It follows that
$$
\begin{array}{lll}
&\displaystyle\int_0^{\infty}\int_0^{\infty}\left|b\mathcal{F}_{q,v}C_{q,\Psi}^{v}(f)(a,b)\right|^{2}\frac{d_qad_qb}{a^2}\\ 
&=\displaystyle\int_0^{\infty}\displaystyle\int_0^{\infty}\left|\sqrt{a}\mathcal{F}_{q,v}\psi(a.)\right|^2\left|b\mathcal{F}_{q,v}f\right|^2\frac{d_qad_qb}{a^2}\\
&=C_{v,\Psi}\displaystyle\int_0^{\infty}\left|b\mathcal{F}_{q,v}f\right|^2d_qb\\
&=C_{v,\Psi}\|\xi\mathcal{F}_{q,v}f(\xi)\|_{q,2,v}^2.
\end{array}
$$
\hfill$\bullet$\\
For the next, we will use the following result which shows a Plancherel/Parseval type rule for the generalized $q$-Bessel wavelet transform (see \cite{Rezguietal}).
\begin{lemma}\label{lemmeRezgui}\cite{Rezguietal}
Let $\Psi$ be a generalized $q$-Bessel wavelet in $\mathcal{L}_{q,2,v}(\widetilde{\mathbb{R}}_{q}^{+})$. Then we have, $\forall f\in \mathcal{L}_{q,2,v}(\widetilde{\mathbb{R}}_{q}^{+})$,
$$
\dfrac{1}{C_{v,\Psi}}\displaystyle\int_{0}^{\infty}\displaystyle\int_{0}^{\infty}|C_{q,\Psi}^{v}(f)(a,b)|^{2}\;b^{2|v|+1}\dfrac{d_{q}a \,d_{q}b}{a^{2}} =\|f\|_{q,2,v}^{2}.
$$
\end{lemma}
\textit{Proof.} We have
$$
\begin{array}{lll}
\medskip&&q^{4|v|+2}\displaystyle\int_{0}^{\infty}\displaystyle\int_{0}^{\infty}|C_{q,\Psi}^{v}(f)(a,b)|^{2}\;b^{2|v|+1}\dfrac{d_{q}a \,d_{q}b}{a^{2}}\\
\medskip&=&q^{4|v|+2}\displaystyle\int_{0}^{\infty}\left(\displaystyle\int_{0}^{\infty}|\mathcal{F}_{q,v}(f)(x)|^{2}|\mathcal{F}_{q,v}(\overline{\Psi_{a}})|^{2}(x)x^{2|v|+1}d_{q}x\right)\dfrac{d_{q}a}{a}\\
\medskip&=&\displaystyle\int_{0}^{\infty}|\mathcal{F}_{q,v}(f)(x)|^{2}\left(|\mathcal{F}_{q,v}(\Psi)(ax)|^{2}\dfrac{d_{q}a}{a}\right)x^{2|v|+1}d_{q}x\\
\medskip&=&C_{v,\Psi}\displaystyle\int_{0}^{\infty}|\mathcal{F}_{q,v}(f)(x)|^{2}x^{2|v|+1}d_{q}x\\
\medskip&=& C_{v,\Psi}\|f\|_{q,2,v}^{2}.
\end{array}
$$
\hfill$\bullet$\\
\textit{Proof of Theorem \ref{theorem1}.} Observe firstly that 
$$
\left\Vert bC_{q,\Psi}^{v}(f)(a,b)\right\Vert_{q,2,v}\left\Vert b\mathcal{F}_{q,v}C_{q,\Psi}^{v}(f)(a,b)\right\Vert_{2}\geq\frac{1}{2}\left\Vert C_{q,\Psi}^{v}(f)(a,b)\right\Vert_{2}^{2}.
$$
Next, it is easy to see that
$$
\left\Vert bC_{q,\Psi}^{v}(f)(a,b)\right\Vert_{q,2,v}=\displaystyle\int_{0}^{\infty}\displaystyle\int_{0}^{\infty}\left|\mathcal{R}f(a,b)\right|^2d_{q}(a,b).
$$
On the other hand, from Lemma \ref{MainResultLemma}, we get
$$
\displaystyle\int_{0}^{\infty}\left|\mathcal{S}f(t)\right|^2\;t^{2|v|+1}d_{q}t=\left\Vert b\mathcal{F}_{q,v}C_{q,\Psi}^{v}(f)(a,b)\right\Vert_{2}.
$$
Therefore, we obtain
$$
\Biggl(\displaystyle\int_{0}^{\infty}\displaystyle\int_{0}^{\infty}\left|\mathcal{R}f(a,b)\right|^2d_{q}(a,b)\Biggr)\Biggl(\displaystyle\int_{0}^{\infty}\left|\mathcal{S}f(t)\right|^2\;t^{2|v|+1}d_{q}t\Biggr)\geq\frac{1}{2}\left\Vert C_{q,\Psi}^{v}(f)(a,b)\right\Vert_{2}^{2}.
$$
So, finally, from Lemma \ref{lemmeRezgui}, we observe that the $q$-wavelet transform is isometric (modulo a normalizing constant, (\cite{Rezguietal})), we get
$$
\Biggl(\displaystyle\int_{0}^{\infty}\displaystyle\int_{0}^{\infty}\left|\mathcal{R}f(a,b)\right|^2d_{q}(a,b)\Biggr)\Biggl(\displaystyle\int_{0}^{\infty}\left|\mathcal{S}f(t)\right|^2\;t^{2|v|+1}d_{q}t\Biggr)\geq K_{q,\Psi}^{v}
\|f\|_{q,2,v}^{2}.
$$
\hfill$\bullet$
\section{Conclusion}
In this paper, an uncertainty principle associated to wavelet transforms in the $q$-calculus framework has been established. A two-parameters extension of the classical Bessel operator is applied to generate a wavelet function which is exploited next to explore a wavelet uncertainty principle already in the $q$-calculus framework. The result joins and extended many works in the same topic such as \cite{Dhaouadi,Manel,Rezguietal}

\end{document}